\documentclass[prl,nofootinbib,superscriptaddress,twocolumn,showpacs]{revtex4-1}

\usepackage{bm}
\usepackage{amsfonts}
\usepackage[pdftex]{color,graphicx}
\usepackage{amssymb}
\usepackage{amsmath}
\usepackage{amsthm}
\usepackage{hyperref}
\usepackage{times}

\newcommand{\ket}[1]{\ensuremath{\left| #1 \right\rangle}}
\newcommand{\bra}[1]{\ensuremath{\left\langle #1 \right|}}

\newcommand{\up}{\uparrow}
\newcommand{\down}{\downarrow}

\DeclareMathOperator{\sgn}{sgn}
\DeclareMathOperator{\Pf}{Pf}

\let\oldmarginpar\marginpar
\renewcommand\marginpar[1]{\-\oldmarginpar[\raggedleft\tiny\color{red} #1]%
{\raggedright\tiny #1}}

\begin{document}


\title{The quantum adiabatic algorithm and scaling of gaps at first order quantum phase transitions}
	
\author{C. R. Laumann}
\affiliation{Department of Physics, Harvard University, Cambridge, MA 02138}

\author{R. Moessner}
\affiliation{Max-Planck-Institut f\"ur Physik komplexer System, 01187 Dresden, Germany}

\author{A. Scardicchio}
\affiliation{Abdus Salam ICTP, and INFN sezione di Trieste, Strada Costiera 11, I-34151, Trieste, Italy}

\author{S.L. Sondhi}
\affiliation{Department of Physics, Princeton University, Princeton, NJ 08544}

\date{\today}

\begin{abstract}
Motivated by the quantum adiabatic algorithm (QAA), we consider the scaling of the Hamiltonian gap at quantum first order transitions, generally expected to be exponentially small in the size of the system. 
However, we show that a quantum antiferromagnetic Ising chain in a staggered field can exhibit a first order transition with  only an {\em algebraically} small gap. 
In addition, we construct a simple classical translationally invariant one-dimensional Hamiltonian containing nearest-neighbour interactions only, which exhibits an exponential gap at a  thermodynamic quantum first-order
transition  of essentially topological origin.
This establishes that (i) the QAA can be successful even across first order transitions but also that (ii) it can fail on exceedingly simple problems  readily solved by inspection, or by classical annealing. 
\end{abstract}

\pacs{03.67.Ac, 05.30.Rt, 75.10.Jm, 75.10.Pq, 75.30.Kz}

\maketitle

The quantum adiabatic algorithm (QAA) \cite{Farhi:2001hy,Farhi:2000tw} promises to harness the power of quantum mechanics, in particular quantum tunneling through energy barriers, in order to solve hard optimization problems more efficiently. Like classical simulated annealing (CSA), part of its appeal is its general purpose ``black box'' nature. Despite the QAA's great potential, the decade since its introduction has seen the discovery of some limitations as a matter of principle. 
The motivation for this work is to understand these limits systematically, clarifying where the algorithm will or will not work. This question is of broad interest: like in the case of CSA, different failure modes point to interesting underlying physics. For instance, critical slowing down and the onset of glassiness are two non-trivial phenomena which can frustrate CSA.

Fundamentally, the QAA fails whenever the  adiabatic evolution encounters an exponentially small Hamiltonian gap. 
It is thus tempting to connect the behavior of the adiabatic algorithm with various kinds of quantum phase transitions, where it is well known that the Hamiltonian gap must close in the thermodynamic limit. A simple heuristic suggests that first order quantum phase transitions are especially problematic: the matrix elements of any local Hamiltonian between macroscopically distinct states will be exponentially small, and hence also the gap of the (barely avoided) crossing.

Here, we analyze some simple one dimensional first order transitions and offer both good news and bad news for the QAA. The good news is that first-order transitions can be accompanied by a finite-size gap which vanishes only algebraically. This is possible because the Hamiltonian gap is not a thermodynamic quantity, and is therefore not necessarily enslaved to the details of the transition, in the same way that a system exhibiting the breaking of a discrete symmetry can exhibit gapless excitations on account of frustrating boundary conditions. 

The bad news is that we add a failure mode to the known limitations of the QAA which we call topological. We construct a simple classical spin model which has near-degenerate ground states which fall into different topological sectors. Adding quantum dynamics prefers the sector with exponentially many ground states, while any degeneracy-lifting interaction favours another containing only O(1) states. The QAA selects the wrong sector in an order-by-disorder mechanism, out of which tunneling becomes exponentially slow as the quantum dynamics is switched off. 

This is remarkable as our example is a translation invariant quasi--one-dimensional Ising model with nearest-neighbour interactions only, the ground state of which is readily found by inspection, CSA or transfer matrix. 
This complements rigorous work on the difficulty of finding the precise ground state energy of translation invariant 1D local Hamiltonians \cite{Gottesman:2009vy}, a task which is QMA$_{\textrm{EXP}}$-complete. Of course, the ground state of our classical Hamiltonian can be found easily, but its strong boundary condition dependence and extreme sensitivity to parameters near the quantum first order transition reflect the features that we believe would arise in any QMA$_{\textrm{EXP}}$-complete simplification of the Hamiltonians considered in \cite{Gottesman:2009vy}.

Before turning to our results, we briefly review known failure modes of the QAA from a physics perspective. 
Quantum first order transitions which provably frustrate the QAA arise in non-local optimization problems whose energy functions do not provide ``basins of attraction'' suitable to local exploration in configuration space 
\cite{Farhi:2010tc,Jorg:2008fj,Farhi:2005uv,vanDam:2001jc}. 
This reflects the inability of local quantum fluctuations to explore non-local landscapes effectively -- either due to extensive disorder \cite{Farhi:2010tc,Jorg:2008fj} or because of golf-course like landscapes with exponentially small holes \cite{Farhi:2005uv,vanDam:2001jc}. 
In models with local energetics, the situation is less clear. 
Thermodynamic calculations within quantum cavity theory \cite{laumann:134424} suggest random quantum first order transitions persist in at least some local models \cite{Jorg:2010ys}, although QMC data is inconclusive \cite{Young:2008bk}. 
Recent controversial work suggests that Anderson localization may arise in configuration space when quantum fluctuations are very weak, leading to `perturbative crossings' and exponentially small gaps \cite{Knysh:2010un,Farhi:2009vh,Altshuler:2010ct,Amin:2009kf}.
Heuristic arguments assuming the presence of `clustering' of pure states in a glassy phase suggest that such crossings may arise throughout an extended regime of the adiabatic evolution \cite{Foini:2010gj}; such crossings may even have been observed in QMC \cite{Young:2010ct}. 
In disordered, geometrically local optimization problems, Griffiths-like effects may arise in which large local regions order before the whole \cite{BenReichardt:2004vd,PhysRevB.51.6411}.

\paragraph{First-order transition with algebraically small gap.} Consider an antiferromagnetic Ising chain in staggered tilted field
\begin{equation}
	\label{eq:af_tfim}
H=J\sum_{\langle ij\rangle} \sigma_i^z\sigma_j^z-\sum_{i} h_i \sigma_i^z - \Gamma\sum_{i} \sigma_i^x
\end{equation}
where $J>0$, $h_i = (-1)^i h$ is a staggered longitudinal field and $\Gamma$ is a uniform transverse field. In the thermodynamic limit at $h=0$, the system exhibits a paramagnetic phase at $\Gamma > J$ and a Ne\'el ordered phase at $\Gamma < J$ where the staggered magnetization $(-1)^i\sigma^z_i$ exhibits long range order. For $\Gamma < J$, as the longitudinal field $h$ is swept down through $0$, the Ne\'el moment $m_h = (-1)^i\langle\sigma^z_i\rangle_h$ jumps from $m_{0+}$ to $m_{0-} = - m_{0+}$ by a finite amount. This corresponds to a first order quantum phase transition where the ground state energy density exhibits a first order cusp (as $L \to \infty$):
\begin{align}
	\frac{1}{L}\partial_h \langle H \rangle = \frac{1}{L} \sum_i (-1)^i \langle \sigma^z_i \rangle = m_h
\end{align}
Each of the staggered phases exhibits a bulk gap to the creation of minority domains. At $h=0$, the single-wall excitation spectrum may be obtained exactly by fermionization, leading to the gapped dispersion $\epsilon(q) = \sqrt{J^2 + \Gamma^2 + 2 J \Gamma \cos{q}}$.

The scaling of the many-body gap at the transition point $h=0$ at finite size $L$ with periodic boundary conditions may be found precisely using fermionization (see Appendix). Here, we provide an intuitive perturbative argument near $\Gamma=0^+$.
For $L$ even, the two degenerate Ne\'el ordered ground states of $H$, $\ket{1} = \ket{\up\down\up\down\cdots\up\down}$ and $\ket{2}=\ket{\down\up\down\up\cdots\down\up}$, each have energy $-L J$. These are separated by a gap $4J$ from $2\binom{L}{2}$ states with a pair of domain walls. The transverse field produces domain walls in pairs and hops individual domain walls. In degenerate perturbation theory, the leading order at which the transverse field couples within the ground state space is $L$ by producing a pair of domain walls and dragging them around the system to annihilate. This leads to an exponentially small gap. 

For $L$ odd, the Ne\'el states do not fit; the lowest energy states must have a broken bond somewhere. This can be in any of $L$ positions and overall Ising symmetry leads to $2L$ degenerate ground states with energy $-J(L-1) + J$:
\begin{align*}
	\ket{0} &= \ket{\vdots \up \down \cdots \up } \quad
	\ket{1} = \ket{\down \vdots \down \cdots \up } \quad
	\cdots \\
	\ket{L} &= \ket{\vdots\down\up \cdots \down} \quad \cdots \quad 
	\ket{2L - 1} = \ket{\up \down \cdots \vdots \down}
\end{align*}
The transverse field acts directly within this state space by hopping the frustrated bond $\vdots$ left or right with amplitude $-\Gamma$. Thus, the effective Hamiltonian in the ground state space is that of a periodic hopping chain of length $2L$:
\begin{align}
	H_{eff} = -\Gamma \sum_{j=0}^{2L-1} \ket{j}\bra{j+1} + \textrm{h.c.}
\end{align}
The spectrum is $\epsilon_k = -2 \Gamma \cos(k)$  with $k = \frac{2\pi}{2L}n$ and $n\in\{0,1,\cdots,2L-1\}$. This leads immediately to a unique ground state energy of $-LJ + 2J - 2\Gamma$ at $k=0$ with a gap to the first excited state of $-2 \Gamma (\cos(\frac{\Pi}{L})-1) \approx \pi^2 / L^2$ which is only algebraically small.

The lesson of this simple calculation is that the Hamiltonian gap is not a thermodynamic quantity, in the same way that an (in)appropriate choice of boundary conditions can force gapless excitations on a system in which only a discrete symmetry is broken. Such non-bulk modes vanish in observables in the thermodynamic limit but modify the many-body gap at finite size.


We next present the topological failure mode of the QAA, for which we first introduce a quantum dimer model, where the topological structure leading to an exponentially small gap is most evident. We then transcribe this result to a system with an unconstrained Hilbert space -- a simple translationally invariant Ising ladder. 
\paragraph{Dimer model.--}
Consider a dimer model on a periodic two leg ladder of length $L$. The dimer Hilbert space is spanned by hardcore dimer
coverings of the ladder. These fall into three sectors which are topological in that they are not connected by any local rearrangement of the dimers. 
The sectors are labelled by a winding number $w$,  the difference between 
the number of dimers on top  and bottom rows (on any fixed plaquette). 
On an even length ladder, there are three sectors $w = \pm 1, 0$, while an odd length ladder is topologically trivial, $w\equiv 0$.
The two `staggered' sectors,  $w = \pm 1$ comprise only one state each while the $w=0$ (`columnar')
sector has $2 F_L + F_{L-1}\sim \tau^{L}$ states, with  $F_L$  the $L$'th Fibonacci
number and $\tau = \frac{1+\sqrt{5}}{2}$  the
golden mean.

\begin{figure}[tbp]
	\centering
		\includegraphics[height=2in]{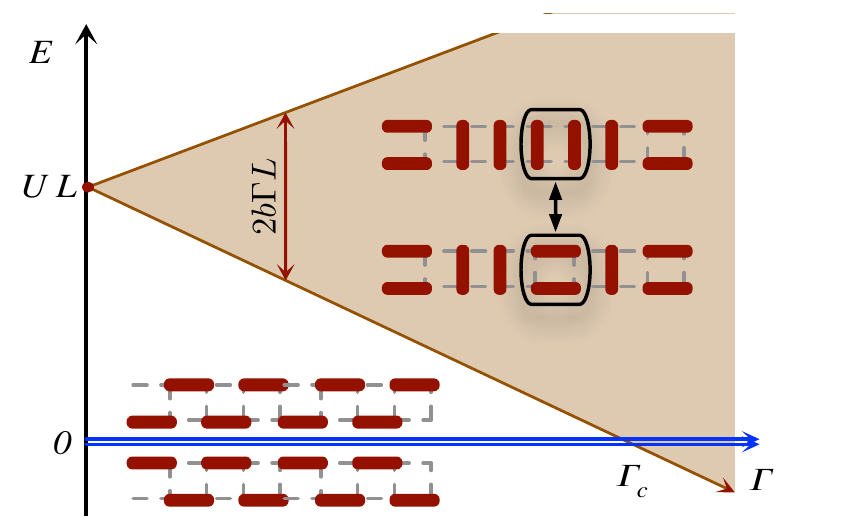}
	\caption{Spectrum of dimer model on even length periodic ladder with dimer configurations overlaid. The two staggered $w=\pm 1$ states are at $E=0$, while the columnar sector $w=0$ splits into a band of resonating dimer states under resonance $\Gamma$.}
	\label{fig:dimer-spec-and-ladders}
\end{figure}

The extraordinarily easy search problem we pose to the QAA is to find the ground state of the Hamiltonian which favours the staggered configurations equally {\em and extensively} over all of the columnar ones. This is encoded by the local, classical Hamiltonian
\begin{align}
	H_{cl} &= U \sum | \, \mathbf{\shortmid}\, \rangle \langle  \,\mathbf{\shortmid}\, | + 2U\sum | = \rangle\langle = | 
\end{align}
assigning extensive energy $U L$ to every state in the $w=0$ sector while leaving the two staggered states $w=\pm 1$ with energy 0. 

It may be surprising that such a simple local Hamiltonian can generate a golf-course energy landscape but it is clear  that such a landscape  is hard for the QAA to maneuver. In fact, the problem is worse than for a golf course, as we explain next. 
 
Any off-diagonal term in the dimer Hilbert space involving only a finite number of rungs in the ladder leaves the winding number $w$ invariant, and hence does not permit transitions between the ground states in different winding sectors. Such off-diagonal terms induce {\em extensive} resonance energies in the entropically dominant columnar sector while leaving the energy of the lonely staggered states completely unperturbed. In other words, not only will switching on a strong quantum perturbation to $H_{cl}$ select 
the wrong state;  switching it off will fail to find the correct sector. 

To make this concrete, consider the simplest quantum resonance term, which flips dimers around a single plaquette:
\begin{align}
	\label{eq:ham_rvb}
	H_{rvb} &= -\Gamma \sum | \, \mathbf{\shortmid} \mathbf{\shortmid}\, \rangle \langle  \,= | + |  \,= \rangle \langle \, \mathbf{\shortmid} \mathbf{\shortmid}\, | \ .
\end{align}
This splits the $w=0$ sector centred at extensive energy $U L$ into a band of width $\sim 2 b L \Gamma$ where $b \approx 0.6$ (from exact diagonalisation numerics, which converges rapidly). The ground state for $w=0$, $\ket{RVB}$, has finite-size energy $E_{RVB} = U L - b_L \Gamma L$, where $b_L$ converges rapidly to $b$ as $L\to \infty$.

At a finite resonance field $\Gamma_c = U / b$, $\ket{RVB}$ undergoes a strict (unavoided) level crossing with the staggered zero energy states (Fig.~\ref{fig:dimer-spec-and-ladders}). In the thermodynamic limit taken through even length periodic chains, this crossing corresponds to a first order quantum phase transition driven by the resonance:
\begin{align}
\partial_\Gamma \epsilon_0 & = \left\{ \begin{array}{lr}
0 & \Gamma < \Gamma_c \\
-b & \Gamma > \Gamma_c
\end{array}\right.
\end{align}

For an odd-length chain, there is only the $w=0$ sector and thus no transition at $\Gamma_c$ ($\partial_\Gamma \epsilon_0 = -b$). This boundary condition dependence of the thermodynamic limit reflects the nonlocality of the dimer Hilbert space.  We address this by transcribing the problem into a  frustrated Ising ladder in a field. 

\begin{figure}[tbp]
	\centering
		\includegraphics[width=0.8\columnwidth]{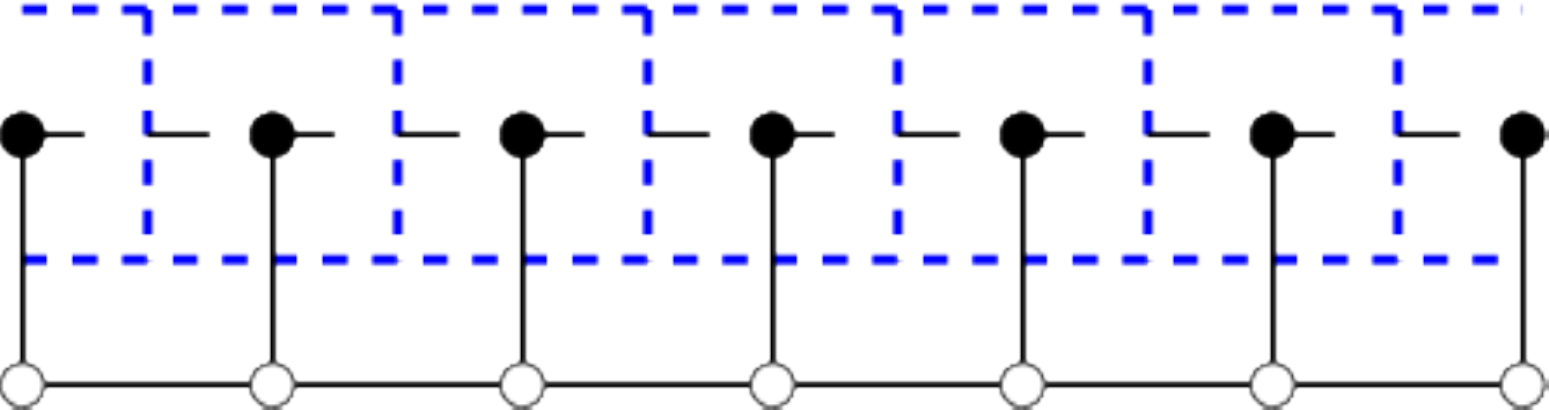}
	\caption{The fully-frustrated  Ising ladder. Location of Ising spins are denoted by filled (empty) circles, subject to fields of strength $-K$ ($U/2$), respectively.  Black solid (dashed) lines represent \mbox{(anti-)ferromagnetic} interactions of strength $K$. Dimers live on the quasi-dual lattice (fat-dashed lines), with a dimer on the top row denoting an unsatisfied field term. 
	The ground states of the $K$ terms are the dimer states, while the $U$ term selects the staggered ones among them. 
	}
	\label{fig:Figures_isingladder}
\end{figure}

\paragraph{Equivalent Ising ladder.}
\label{sec:tll}
The dimer model on the two-leg ladder can be turned, via  a duality transformation, into a frustrated Ising ladder, the ground states of which map onto the dimer states \cite{Moessner:2000td,Moessner:2001da}. The simplest Ising model which effectively reproduces the dimer model physics described above turns out to be a two-leg ladder, which features an external field and {\em nearest-neighbour interactions only}, see Fig.~\ref{fig:Figures_isingladder}. The corresponding classical Hamiltonian is
\begin{align}
	\label{eq:ising-ladder}
	H_I = -\sum_{<ij>} K_{ij} \sigma^z_i \sigma^z_j - K \sum_{i\textrm{ upper}} \sigma^z_i + \frac{U}{2} \sum_{i\textrm{ lower}} \sigma^z_i
\end{align}
where $K_{ij} = (-)K$ for solid (dashed) links $ij$, respectively. 

The ladder is constructed such that (at least) one term of order $K$ (spin interaction or field in the top row) per square must remain unsatisfied; denoting this term by a dimer placed on a fat-dashed line yields the dimer states discussed above. 
Ising configurations with extra dimers are \emph{defective}; they have higher energy in units of $K$.
We observe that non-defective staggered dimer configurations correspond to all bottom row spins negative. 
These spins satisfy the $U$-field, while those in the $w=0$ configurations do not. Thus, $H_I$ encodes the same low energy dimer energetics as $H_{cl}$ in the pure dimer model.

Local quantum dynamics with matrix elements within the flippable $(w=0)$ sector, such as a transverse field $\Gamma \sum \sigma^x$,  lead to the order-by-disorder selection discussed above. This favors a resonating state in the $w=0$ sector over the rigid staggered states favored by $U$.  
As the Hilbert space of this model is now (local and) larger than for the dimer model, the excited state spectrum is richer and the transition more complicated. 
The presence of local topological defects at energy $O(K)$ breaks the strict conservation of $w$ seen in the dimer model, lifting the level crossings. However, for large $K/U$, the first order transition seen in the pure dimer model persists,  acquiring an exponentially small gap due to the virtual hopping of defects around the system.

\begin{figure}[tbp]
	\centering
		\includegraphics{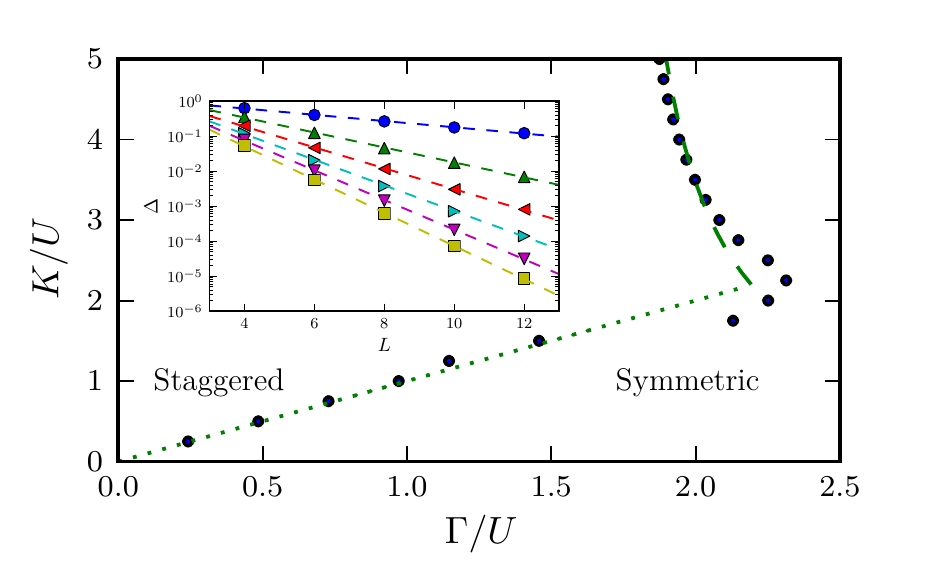}
	\caption{Phase diagram of frustrated two leg ladder. Points indicate position of minimum gap in $k=0$ sector as a function of $\Gamma$ at fixed $K$, $L=12$. Dotted line $\Gamma = K$ indicates second order condensation predicted to leading order in $K/U$; dashed line indicates first order transition predicted to second order in $U/K$ (see text). (inset) Exponential scaling of the gap for $K$ in the first order transition regime. Dashed lines are best fit exponentials to minimum gap data. From top to bottom, $K/U = 2.5 -5.0$ in steps of $0.5$.}
	\label{fig:Figures_pd-mingap-comb}
\end{figure}

At large $K\gg U$, the pure dimer states are energetically well separated from the states with defects. To leading order for the pure dimer $w=0$ sector, the transverse field $\Gamma$ on the top row acts exactly as $H_{rvb}$ of Eq.~\eqref{eq:ham_rvb}, while the transverse field on the bottom row produces defects. Thus, for infinite $K$ and even $L$, we precisely recover the dimer model and its first order transition at $\Gamma_c = U/b$. At large but finite $K$, the transition is shifted by the resonance energy associated with virtual defect states. Thus, to second order, the first order transition curve shifts to $\Gamma_c = U/b + U^2/4Kb^3$, in agreement  with the exact diagonalization results (see Fig.~\ref{fig:Figures_pd-mingap-comb}). 
For odd lengths, the first order transition persists although the finite size effects are more complicated; nonetheless, the minimum gap arises due to the  production and dragging of defects around the system and it remains exponentially small.

For $K \ll U$, the transition from staggered to translation invariant  phase persists as a second order condensation transition. At $U \gg K, \Gamma$, the $U$ field pins the bottom row of spins to be negative, leading to an effective Hamiltonian for the top row of an antiferromagnetic Ising chain with couplings $K$ in a transverse field $\Gamma$ and longitudinal field $h_{eff} = K(1+\langle \sigma^z_{bot}\rangle) \approx 2 K \Gamma^2 / U^2$ at leading order. This chain indeed exhibits the usual second order transition at $\Gamma \approx K - O(h_{eff}^2/K)= K -O(K \Gamma^4 / U^4)$ \cite{Ovchinnikov:2003eh}. In dimer language, the liquid state  at low $K$ corresponds to a condensate of defects while that at higher $K$ is a resonating dimer liquid.

\paragraph{Exact diagonalisation of the Ising model}
We back up the above assertions  with  a  detailed study of the two leg Ising ladder, for which we describe an interesting phase diagram. We exactly diagonalize the zero momentum sector at even lengths $L$ from 4 to 12 ($U=1$). The points in the phase diagram in Fig.~\ref{fig:Figures_pd-mingap-comb} are determined by the $\Gamma$ corresponding to the minimum gap at fixed $K$ at size $L=12$. 
Away from the multicritical point where the first order and second order transition curves meet, this estimate is indistinguishable from the estimate made by finite size scaling estimates; near the meeting point such estimates are suspect at these small sizes.

Viewing the two leg ladder Hamiltonian as a optimization problem for the QAA, the important feature is the scaling of the minimum gap $\Delta$ as a function of $L$. These are plotted in the inset to Fig.~\ref{fig:Figures_pd-mingap-comb} for $K \ge 2.5$ where the transition lies on the first order curve. As expected, the exponential fit to $\Delta(L)$ is very good, exhibiting a decay constant growing linearly with $\log(K/\Gamma)$ for $K$ up to $8$ (fit not shown).

\paragraph{Algorithmic implications.}

It is instructive to consider the behavior of the classic ``black box''
optimization algorithm: classical simulated annealing (CSA) with local spin
flip dynamics. Similar to the quantum order-by-disorder effect, high temperature
favors the entropy of the columnar sector, which freezes out at low temperatures
due to the energetic preference for the staggered sector. Unlike the QAA,
however, CSA may dissipate energy, so that defects in the low temperature regime
anneal out diffusively with diffusion constants depending on temperature but not
system size. This leads to $O(L^3)$ upper bounds on the time needed to find the
ground state.

Hence, CSA finds a ground state in the simplest incarnation. There are ways to fix the QAA as well, but these appear to require insight into the nature of the solution and are in spirit equivalent to finding the solution by knowing something about it, \emph{i.e.}\ solving the problem by inspection. 


\paragraph{Conclusions.--}

The quantum adiabatic algorithm holds much promise as a practical tool for solving hard optimization problems, and as such has suffered a barrage of theoretical attacks over the last decade. Here, we contribute two simple cautionary tales to its analysis: first, that the finite size scaling of the Hamiltonian gap exhibited at  thermodynamic quantum phase transitions must be treated very carefully as it may be exponentially sensitive to non-thermodynamic details and second, that topology and entropy can overwhelm local quantum dynamics, even in translation invariant, local qubit systems. 
Our results also raise the intriguing possibility of a further classification of quantum first order transitions by non-thermodynamic criteria such as the sensitivity to boundary conditions of their Hamiltonian gaps. 


\paragraph{Acknowledgments.--}
We thank E. Farhi, S. Gutman, J. Goldstone, D. Gossett and P. Young for discussions. S.L.S. was supported by NSF grant number PHY-1005429. C.R.L. acknowledges support from a Lawrence Gollub fellowship and the NSF through a grant for ITAMP at Harvard University. A.S.\ thanks the CTP at MIT for hospitality and financial support.


\bibliography{dimerladder}

\appendix

\section{Many-body gaps in quantum Ising chain by fermionization} 
\label{sec:fermionic_tfim}

We consider the quantum Ising chain as in the main text, but using a rotated basis of Pauli operators ($\sigma^x \leftrightarrow \sigma^z$) to connect with the convention of \cite{Kitaev:2010uq}:
\begin{align}
	\label{eq:tfim_xxz}
	H = -J\sum_{\langle ij\rangle} \sigma_i^x\sigma_j^x - \Gamma\sum_{i} \sigma_i^z
\end{align}
We define the Majorana Jordan-Wigner transformation:
\begin{align*}
	\sigma_i^x &= \prod_{j<i} (-i \gamma_j \gamma_j') \gamma_i\\
	\sigma_i^y &= \prod_{j<i} (-i \gamma_j \gamma_J') \gamma_i'\\
	\sigma_i^z &= -i \gamma_i \gamma_i'
\end{align*}
where the $2L$ operators $\gamma_i, \gamma_i'$ are Majorana fermion operators satisfying the standard anticommutation relations:
\begin{align*}
	\{\gamma_i, \gamma_j\} = 2 \delta_{ij} \qquad 
		\{\gamma_i', \gamma_j'\} = 2 \delta_{ij} \qquad
	\{\gamma_i, \gamma_j'\} = 0
\end{align*}
Finally, define the string operators
\begin{align}
	S_i = \prod_{j<i} (-i \gamma_j \gamma_j') = \prod_{j<i} \sigma^z_j
\end{align}
and the total parity operator
\begin{align}
	\hat{P} = S_{L+1} = \prod_{j} (-i \gamma_j \gamma_j') = \prod_{j} \sigma^z_j
\end{align}
The parity operator $\hat{P}$ implements the global Ising symmetry of the model.

On a periodic chain of length $L$, the Hamiltonian may be rewritten in the fermionic language
\begin{align*}
	H &= -J \sum_{j=1}^{L-1} S_j \gamma_j S_{j+1} \gamma_{j+1} - J S_L \gamma_L \gamma_1 - \Gamma \sum_j (-i \gamma_j \gamma_j') \\
	&= J i \sum_{j=1}^{L-1} \gamma_j' \gamma_{j+1} - J \hat{P} i \gamma_L' \gamma_1 + \Gamma i \sum_j \gamma_j \gamma_j' 
\end{align*}
Within each parity sector, the Hamiltonian is quadratic in Majorana fermions and corresponds to a bipartite hopping chain with a two site unit cell and either periodic ($P=-1$) or anti-periodic ($P=1$) boundary conditions. We may diagonalize such a problem by Fourier transform to find (positive energy modes):
\begin{align}
	\epsilon(k) = \sqrt{J^2 + \Gamma^2 - 2 J \Gamma \cos(k)}
\end{align}
where we take $k = \frac{2\pi}{L}n$ for $P=-1$ and $k = \frac{2\pi}{L} (n+1/2)$ for $P = 1$, $n=0\cdots L -1$. The vacuum state for the quadratic problem $H_\pm$ with $P=\pm1$ is $\ket{\Omega, \pm}$ and has energy
\begin{align}\label{eq:vac_energy}
	H_\pm \ket{\Omega, \pm} = - \sum_k \epsilon(k) \ket{\Omega, \pm} = E_\pm^0 \ket{\Omega,\pm}
\end{align}
where $k$ runs over the appropriate momenta for the boundary condition. 

In Eq.~\eqref{eq:vac_energy} we have calculated the energy of $\ket{\Omega,\pm}$ within the effective Hamiltonian $H_\pm$ which fixes the value $P=\pm1$, ignoring the consistency condition that $\ket{\Omega,\pm}$ actually have parity $P$. For a general quadratic Majorana Hamiltonian, 
\begin{align}
	H = \frac{i}{4} \gamma_i h_{ij} \gamma_j
\end{align}
where $h_{ij}$ is a real antisymmetric coefficient matrix and $i,j$ run over $1,1',2,2'\cdots$, the lowest energy state $\ket{\Omega}$ has parity
\begin{align}
	\hat{P} \ket{\Omega} = \sgn(\Pf(h)) \ket{\Omega}.
\end{align}
If $\hat{P}\ket{\Omega,\pm} = \pm \ket{\Omega, \pm}$, the vacuum energy corresponds to the ground state energy in that parity sector. If not, the ground state in the sector will correspond to $c^\dagger_q \ket{\Omega,\pm}$ where $c^\dagger_q$ creates the lowest energy excitation of the quadratic Hamiltonian as this is the lowest state with the correct parity.

To compute the parity of the vacua for the magnetically ordered phase, we exploit the adiabatic invariance of $\sgn(\Pf(h))$. Thus, we may work at the trivial points in the phase diagram where either $J$ or $\Gamma$ is $0$. First, we observe that for $J=0, \Gamma=1$ in the paramagnetic phase, the Hamiltonian reduces to
\begin{align}
	H = i\sum \gamma_j \gamma_j'
\end{align}
whose ground state clearly sets $-i \gamma_j \gamma_j' = 1$ for all $j$. Thus the parity of the vacuum is $1$ for both boundary conditions and the system has a unique ground state in the $P=1$ sector with a gap to single particle excitations in the $P=-1$ sector as expected.

Now, we consider the ferromagnetic $J=1, \Gamma=0$ Hamiltonian:
\begin{equation}
	H = i \sum \gamma_j' \gamma_{j+1} - P i \gamma_L' \gamma_1
\end{equation}
For $P=-1$, $h_{FM}$ for this Hamiltonian is precisely a translation by 1 step in the length $2L$ chain of Majoranas. Thus, $h_{FM} = R h_{PM} R^T$ where $R$ translates every site by 1. Since $\Pf(R h_{PM} R^T) = \det(R) \Pf(h_{PM})$, and $\det(R) = (-1)^{2L-1} = -1$, we have that $P \ket{\Omega,-} = - \ket{\Omega,-}$. Similarly, flipping $P$ to $+1$  multiplies one row and column of $h_{FM}$ by -1, flipping $\sgn(\Pf(h_{FM}))$. Thus, $P\ket{\Omega,\pm} = \pm \ket{\Omega,\pm}$ in the ferromagnetic phase and therefore the ground state of each sector is the vacuum state for the effective quadratic Hamiltonian in that sector. Thus, for the ferromagnet, we have the bottom two states have energy:
\begin{align*}
	E_+^0 &= -\sum_n \epsilon(\frac{2\pi}{L}(n+1/2)) \\
	E_-^0 &= -\sum_n \epsilon(\frac{2\pi}{L}(n))
\end{align*}
Since $\epsilon(k)$ is analytic and lives on the circle $k \in (0,2\pi)$, the Fourier series coefficients $\tilde\epsilon_m$ decay faster than any power law in $m$. Using this, it is straightforward to show that the $E_+^0-E_-^0$ is smaller than any power law in $L$.

Finally, we turn to antiferromagnetic $J=-1, \Gamma=0$. In this case,
\begin{equation}
	h_{AFM} = - h_{FM}
\end{equation}
and thus $\Pf(h_{AFM}) = (-1)^L \Pf(h_{FM})$. For $L$ even, this means that there are two ground states identical to those found in the ferromagnetic case with an exponentially small separation in $L$. For $L$ odd, the parity of the vacuum in both sectors is wrong:
\begin{align}
	\hat{P}\ket{\Omega,\pm} = \mp \ket{\Omega,\pm}
\end{align}
and thus the two lowest energy states correspond to populating the lowest energy excitation in each sector:
\begin{align}
	E_+ &= E_+^0 + 2 \epsilon(\pi + \frac{\pi}{L}) \\
	E_- &= E_-^0 + 2 \epsilon(\pi)
\end{align}
Expanding the single particle dispersion $\epsilon$ at its quadratic minimum at $k=\pi$, this leads to a power law small gap
\begin{align}
	E_+ - E_- & \approx 2 \Gamma \frac{\pi^2}{L^2}.
\end{align}

\end{document}